\newcommand{\q}{\vec q}
\newcommand{\x}{\vec x}

\newcommand{\xp}{\vec x^{\, \prime}}
\newcommand{\xpp}{\vec x^{\,\prime\prime }}
\newcommand{\xppp}{\vec x^{\, \prime\prime\prime }}

\newcommand{\im}{{\rm  i}}
\newcommand{\lcoh}{{l_{\rm  coh}}}

\documentclass[pra,aps,epsfig,showpacs,amsmath,multicol]{revtex4}

\usepackage{graphics}
\begin{document}
\title{Correlated imaging, quantum and classical}
\author{A.~Gatti, E.~Brambilla, M. Bache and L.~A.~Lugiato}
\address{INFM, Dipartimento di Scienze CC.FF.MM.,
Universit\`a dell'Insubria, Via Valleggio 11, 22100 Como, Italy}
\begin{abstract}
  We analytically show that it is possible to perform coherent imaging
  by using the classical correlation of two beams obtained by
  splitting incoherent thermal radiation.  A formal analogy is
  demonstrated between two such classically correlated beams and two
  entangled beams produced by parametric down-conversion.  Because of
  this analogy, the classical beams can mimic qualitatively all the
  imaging properties of the entangled beams, even in ways which up to
  now were not believed possible.  A key feature is that these
  classical beams are spatially correlated both in the near-field and
  in the far-field.  Using realistic numerical simulations the
  performances of a quasi-thermal and a parametric down-conversion
  source are shown to be closely similar, both for what concerns the
  resolution and statistical properties.
  The results of this paper provide a new scenario for the discussion
  of what role the entanglement plays in correlated imaging.
\end{abstract}
\centerline{Version \today}
\maketitle

\section{Introduction}
The topic of entangled imaging has attracted noteworthy attention in
recent years \cite{bib1,bib2,bib3,bib4,bib5,bib6,bib7,bib13,bib14}.
This technique exploits the quantum entanglement of the state
generated by parametric down-conversion (PDC), in order to retrieve
information about an unknown object.  In the regime of single photon-pair 
production of PDC, the photons of a pair are spatially separated
and each propagates through a distinct imaging system, usually called
the test and the reference arm.  An object is located in the test arm.
Information about the spatial distribution of the object is obtained
by registering the coincidence counts as a function of the transverse position of
the photon in the reference arm, which holds a known reference system
\cite{bib1,bib2,bib3,bib4,bib5,bib6}.  In the regime of a large number
of photon pairs, this procedure is generalized to the measurement of
the signal-idler spatial correlation function of intensity
fluctuations \cite{bib7}. Such a two-arm configuration provides more
flexibility in comparison with standard imaging procedures. For
example, there is the possibility of illuminating the object at a
given light frequency in the test arm and of performing a spatially
resolved detection in the other arm with a different light frequency,
or of processing the information from the object by only operating on
the imaging system of the reference arm \cite{bib6}.  In addition, it
opens the possibility for performing coherent imaging by using, in a
sense, spatially incoherent light, since each of the two
down-converted beams taken separately is described by a thermal-like
mixture and only the two-beam state
is pure (see \cite{bib5} and \cite{bib7}).

\begin{figure}[h]
{\scalebox{.35}{\includegraphics*{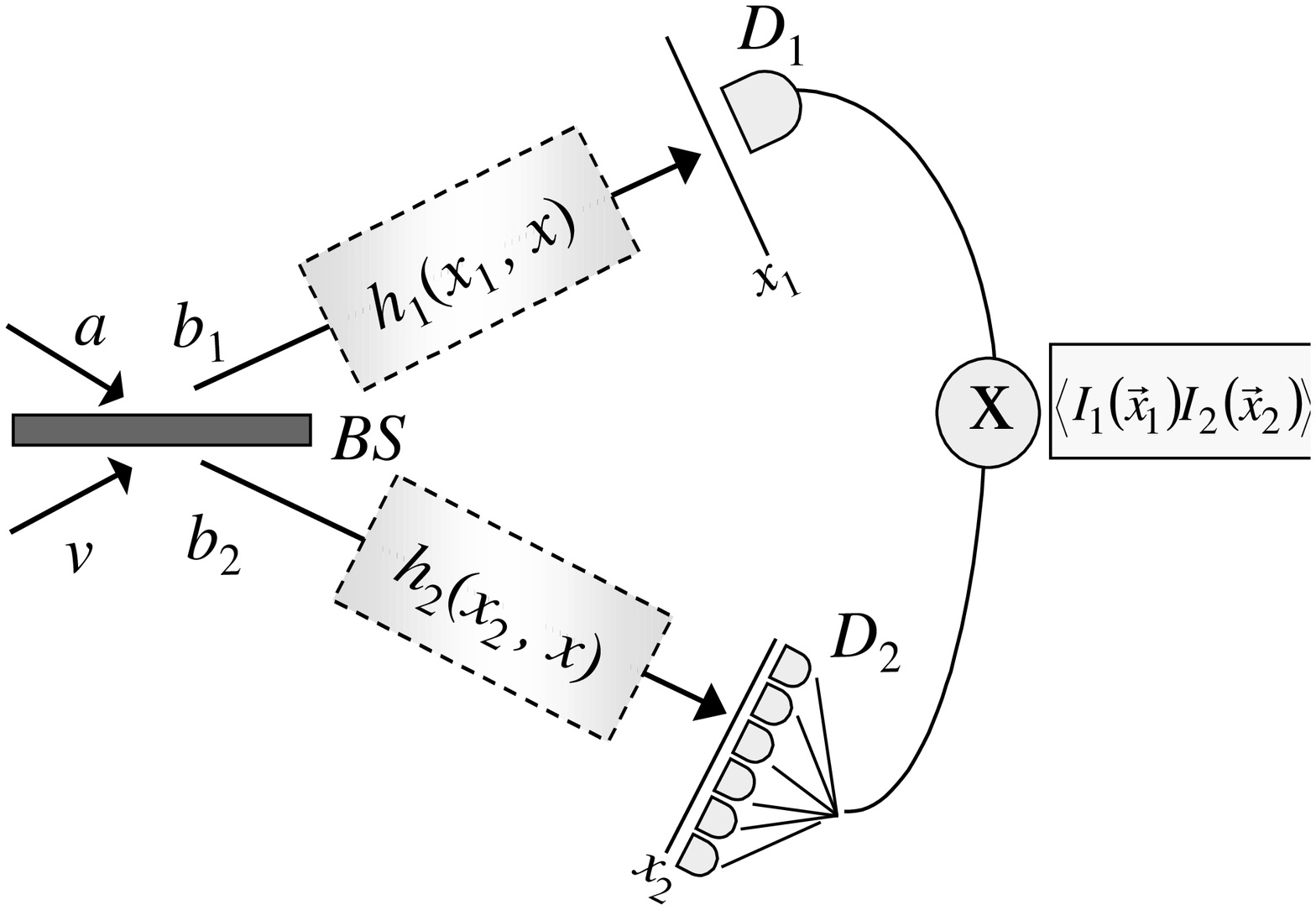}}}
\caption{Correlated imaging with incoherent thermal light. The thermal
  beam $a$ at the beam-splitter $BS$ is divided into two beams, $b_1$
  and $b_2$, which travel through respectively a test and a reference
  system, described by their impulse response functions $h_1$ and
  $h_2$. The test arm 1 includes an object. Detector
  $D_1$ is either a point-like detector or a bucket detector. $D_2$ is
  an array of pixel detectors.  $v$ is a vacuum field.}
\label{fig2}
\end{figure}

In this paper we show that such a scheme can be implemented using
truly incoherent light, as the radiation produced by a thermal (or
quasi-thermal) source.  A comparison between thermal and photon-pair
emission was performed in \cite{bib8b}, where an underlying duality
accompanies the mathematical similarity between the two cases.  Here
we consider a different scheme (Fig.~\ref{fig2}) appropriate for
correlated imaging, in which a thermal beam is divided by a
beam-splitter (BS) and the two outgoing beams are handled in the same
way as the PDC beams in entangled imaging. Our analysis points out a
precise formal analogy between the PDC and the thermal case. This
analogy opens the possibility for using classically correlated thermal
light for correlated imaging in the same way as entangled beams from
PDC.
Currently there is a very lively debate whether quantum entanglement
is a necessary ingredient to perform correlated imaging
\cite{bib5,bib13,bib7,bib14}. The first coincidence imaging
experiments were performed using entangled photons from PDC
\cite{bib3,bib4}. At that time, the Authors of Ref. \cite{bib3}
suggested that {``it is possible to imagine some type of classical
source that could partially emulate this behavior".
A more recent theoretical analysis \cite{bib5} gave arguments that
``the distributed quantum-imaging scheme truly requires entanglement
in the source and cannot be achieved using a classical source with
correlations but without entanglement".
The topic became hot after the ghost image experiment of Ref.
\cite{bib3} was successfully reproduced using classically correlated
beams \cite{bib13}. In this experiment a classical source produced
pairs of single-mode angularly correlated pulses that served as
classical analogs of momentum correlated pairs of photons produced by
PDC.  In the accompanying theoretical discussion, the Authors
presented arguments that although the results of any single experiment
in quantum imaging could be reproduced by classical sources with
proper statistical correlation, a given classical source cannot
emulate the behaviour of a quantum entangled source for any arbitrary
test and reference systems.  In agreement with this we showed in a
recent theoretical Letter \cite{bib7} that on the one hand the results
of any single quantum-imaging experiment performed with entangled
beams in a pure state could be exactly reproduced by using separable
(i.e.  non-entangled) states.
On the other hand, we showed that a key feature of the entangled state produced by PDC
is the simultaneous presence of  spatial correlation at the quantum level in the
near field and the far field of the signal-idler beams (see also
\cite{bib8}). This corresponds to the simultaneous quantum correlation
of position and momentum of the photons in each pair. 
We showed that this feature could be exploited to produce both the image and
the diffraction pattern of an object
by solely operating on the reference arm, and argued that this could
be done only in the presence of quantum entanglement.  This
interpretation was received rather well in the quantum imaging
community and was generally viewed as a possibility to discriminate
between the presence of quantum entanglement and classical correlation
in the source.


In this paper we will analyse a counter-example, which partially
contradicts the picture emerging from Refs. \cite{bib7,bib13}. The
classical, thermal beams created by splitting thermal radiation 
have several features that
distinguish them from other non-entangled beams considered in the previous
literature 
\cite{bib7,bib13}. First
of all, they 
are spatially multi-mode, unlike those considered in Ref.  \cite{bib13}.
Second, as we will see in Section \ref{comparison} they have an
imperfect level of spatial correlation, limited by the classical
shot-noise introduced by the beam-splitter.  Nonetheless, they exhibit a  
high level of spatial correlation in both the near-field
and far-field planes, in contrast to the separable mixtures considered in
Ref. \cite{bib7} that were perfectly correlated in either plane.
We will show that this imperfect spatial correlation in both planes,
although being completely classical, is enough to qualitatively
reproduce {\em all} the features of the entangled imaging, provided
that the spatial coherence properties of the thermal source are
properly engineered.  Finally, they are probably the best classical
analogs of the entangled signal-idler beams produced by PDC, since the
marginal statistics of the signal or idler beam alone is a thermal
statistics. Thus, they should provide an optimal test bed for
understanding the role of entanglement in correlated imaging.

In Section II we demonstrate theoretically the analogy between thermal
and entangled PDC beams in correlated imaging.  Section III discusses
in analytical terms a specific imaging scheme.
In Sec. IV we discuss the origin of the spatial correlation in the
thermal case and relate it to the entangled case. In
Sec.~\ref{sec:Imag-perf-classical} the performances of the two cases
are compared and we show a key numerical example that drops the
spatio-temporal translational invariance assumed in the analytical
treatment.  Section VI contains the conclusions and a discussion.
Finally, the Appendix discusses the degree and visibility of the
correlation between the thermal beams when the finite detection area of the
measurement apparatus is taken into account.

\section{Analogy between thermal and entangled beams in correlated imaging}
In this section we are going to show a close analogy between the use
of thermal light and entangled beams from PDC in the imaging schemes
based on correlation measurements.  For the sake of comparison, the
two cases will be treated in parallel. In the analytical treatment we
consider for simplicity only spatial variables and ignore the time
argument, which corresponds to using a narrow frequency filter.  We
will come back to this point in Sec.~\ref{sec:Imag-perf-classical}.
In addition, we assume translational invariance in the transverse
plane, which amounts to requiring that the cross-section of the source
is much larger than the object and all the optical elements. 

In the entangled case, the signal and idler fields are generated in a
type II $\chi^{(2)}$ crystal by a PDC process. Our starting point is
the input-output relations of the crystal, which in the plane-wave
pump approximation read \cite{bib7,bib8,bib9}
\begin{equation}
b_i (\q)= U_i(\q) a_i (\q) + V_i (\q) a_j^\dagger (-\q) \quad i \ne j =1,2 \, .
                    \label{eq1}
\end{equation}
Here, $b_i (\q) = \int \frac{{\rm d} \x}{2 \pi} e^{-\im \q\cdot \x}
b_i (\x) $, where $b_i(\x)$ are the signal $(i=1)$ and idler $(i=2)$
field envelope operators at the output face of the crystal
(distinguished by their orthogonal polarizations), $\x$ being the
position in the transverse plane. $ a_i \, , \; i=1,2 $ are the
corresponding fields at the input face of the crystal, and are taken
to be in the vacuum state.  The gain functions $U_i, V_i$ are for
example given in \cite{bib8}.

In the thermal case, we start from the input-output relations of a
beam splitter
\begin{eqnarray}
b_1 (\x )= r a(\x) + t v (\x) \, ,\quad
b_2 (\x )= t a (\x)+ r v(\x) 
                                        \label{BS} \: ,
\end{eqnarray}
where $t$ and $r$ are the complex transmission and reflection
coefficients of the mirror, $a$ is a thermal field and $v$ is a vacuum
field uncorrelated from $a$. We assume that the thermal state $a(\x)$
is characterized by a Gaussian field statistics, in which any
correlation function of arbitrary order is expressed via the second
order correlation function \cite{bib10}:
\begin{eqnarray}
\Gamma (\x, \xp) &=& \langle a^\dagger (\x) a (\xp) \rangle \nonumber \\
&=& \int  \frac{{\rm d} \q}{(2 \pi)^2}  e^{-\im \q\cdot (\x-\xp)} \langle n (\q) \rangle_{\rm th}
\label{gamma} \: .
\end{eqnarray}
Here $\langle n (\q) \rangle_{\rm th} $ denotes the expectation value of
the photon number in mode $\q$ in the thermal state. In writing the
second line of this equation, we implicitly used the hypothesis of
translational invariance of the source, under which $\Gamma (\x,\xp)=
\Gamma (\x-\xp)$. In particular, the following factorization property
holds \cite{bib10}:
\begin{eqnarray}
 \langle : a^\dagger (\x) a (\xp) &&a^\dagger (\xpp) a (\xppp) : \rangle
 =
\langle  a^\dagger (\x) a (\xp) \rangle \langle  a^\dagger (\xpp) a
(\xppp) \rangle  
+
\langle \, a^\dagger (\x) a (\xppp) \rangle \langle a^\dagger (\xpp) a (\xp) \,  \rangle \; ,
\label{gamma2}
\end{eqnarray}
where $ : \, : $ indicates normal ordering.

In both the PDC and the thermal case each of the two outgoing beams
travels through a distinct imaging system, described by its impulse
response functions $h_1 (\x_1,\xp_1)$ and $h_2 (\x_2,\xp_2)$,
respectively (see Fig.~\ref{fig2}).  Arm 1 includes an object. Beam 1
is detected by $D_1$, which is either a point-like detector or by a
``bucket" detector which collects all the light in the detection plane
\cite{bib6}; in any case $D_1$ gives no information on the object
spatial distribution. In the other arm the detector $D_2$ spatially
resolves the light fluctuations, as for example an array of pixel
detectors.  The fields at the detection planes are given by
\begin{equation}
c_i (\x_i) = \int {\rm d} \xp_i h_i( \x_i,\x_i\,') b_i (\x_i\,') + L_i (\x_i)  \quad i=1,2 \, ,
                \label{cfields}
\end{equation}
where $L_1, L_2$ account for possible losses in the imaging systems,
and depend on vacuum field operators uncorrelated from $b_1, b_2$. 
Information about the object is extracted by
measuring the spatial correlation function of the intensities detected by $D_1$ and $D_2$,
as a function of the position $\x_2$ of the pixel of $D_2$:
\begin{equation}
\langle I_1 (\x_1) I_2 (\x_2) \rangle =
\langle  c_1^\dagger (\x_1) c_1 (\x_1) c_2^\dagger (\x_2) c_2 (\x_2) \rangle \; .
\label{eq5}
\end{equation}
All the object information
is concentrated in the correlation function of intensity fluctuations:
\begin{equation}
G(\x_1, \x_2) = \langle I_1 (\x_1) I_2 (\x_2) \rangle - \langle I_1 (\x_1)\rangle \langle I_2 (\x_2) \rangle \; ,
\label{eq6}
\end{equation}
where $\langle I_i (\x_i)\rangle = \langle c_i^\dagger(\x_i) c_i(\x_i)
\rangle $ is the mean intensity of the $i$-th beam. When using a
bucket detector in arm 1, the measured quantity corresponds to the
integral over $\x_1$ of both sides of Eq.~(\ref{eq6}). Since $c_1$ and
$c_2^\dagger$ commute, all the terms in Eqs.~(\ref{eq5}),(\ref{eq6})
are normally ordered and $L_1, L_2$ can be neglected, thus obtaining
\begin{eqnarray}
 G(\x_1, \x_2) =&& \int {\rm d} \xp_1
\int {\rm d} \xpp_1
\int {\rm d} \xp_2 \int {\rm d} \xpp_2 
  h_1^*(\x_1, \xpp_1) h_1 (\x_1, \xp_1) h_2^* (\x_2, \xpp_2) h_2
  (\x_2, \xp_2) 
\nonumber \\ &&
\Big[
\langle  b_1^\dagger (\xpp_1) b_1 (\xp_1) b_2^\dagger (\xpp_2) b_2
(\xp_2) \rangle 
- 
\langle  b_1^\dagger (\xpp_1) b_1 (\xp_1)\rangle \, \langle
b_2^\dagger (\xpp_2) b_2 (\xp_2) \rangle \Big] \: .
\label{eq7}
\end{eqnarray}

In the thermal case, by taking into account the transformation (\ref{BS})
and that $v$ is in the vacuum state, $b_1$  and $b_2$  in Eq.~(\ref{eq7}) can be simply replaced
by $r a$ and $ta$, respectively. Next, by using Eq.~(\ref{gamma2}), we arrive at the final result
\begin{equation}
G_{\rm th}  (\x_1, \x_2) = |tr|^2\left| \int {\rm d} \xp_1
\int {\rm d} \xp_2  h_1^* (\x_1, \xp_1) h_2 (\x_2, \xp_2) \langle a^\dagger (\xp_1) a(\xp_2) \rangle
\right|^2 \: ,
\label{eq12}
\end{equation}
where $\langle a^\dagger (\xp_1) a(\xp_2) \rangle$ is given by
(\ref{gamma}). 

Similarly, the four-point correlation function in Eq.~(\ref{eq7}) has
special factorization properties also in the PDC case.  As it can be
obtained from Eq.~(\ref{eq1}) (see also \cite{bib8}),
\begin{eqnarray}
\langle  b_1^\dagger (\xpp_1) b_1 (\xp_1)  b_2^\dagger (\xpp_2) b_2 (\xp_2) \rangle
&=&
\langle  b_1^\dagger (\xpp_1) b_1 (\xp_1)\rangle \, \langle b_2^\dagger (\xpp_2) b_2 (\xp_2) \rangle \nonumber \\
&+&
\langle  b_1^\dagger (\xpp_1) b_2^\dagger (\xpp_2)\rangle \, \langle b_1 (\xp_1) b_2(\xp_2) \rangle
\: .
\label{eq8}
\end{eqnarray}
By inserting this result in Eq. (\ref{eq7}), one obtains
\begin{equation}
G_{\rm PDC}(\x_1, \x_2)  = \left| \int {\rm d} \xp_1
\int {\rm d} \xp_2  h_1 (\x_1, \xp_1) h_2 (\x_2, \xp_2) \langle b_1 (\xp_1) b_2(\xp_2) \rangle
\right|^2 \: ,
\label{eq9}
\end{equation}
where by using relations (\ref{eq1})
\begin{equation}
\langle b_1 (\xp_1) b_2(\xp_2) \rangle
=
\int \frac{ {\rm d} \q }{(2\pi)^2} e^{\im \q \cdot (\xp_1 -\xp_2)} U_1 (\q) V_2 (-\q)
\: .
\label{gammaentangled}
\end{equation}

At this point the analogy between the results in the two cases clearly
emerges.  
Apart from the numerical factor $|tr|^2$ and the presence of $h_1^*$
instead of $h_1$, the thermal second-order correlation $\langle
a^\dagger (\x) a(\xp) \rangle $ in Eq.~(\ref{eq12}) plays the same
role as the PDC signal-idler correlation function $\langle b_1 (\x)
b_2(\xp) \rangle $ in Eq.~(\ref{eq9}). Consequently from
Eqs.~(\ref{gamma}) and~(\ref{gammaentangled}), the thermal mean photon
number $\langle n(\q) \rangle_{\rm th}$ plays the same role as $U_1(\q)V_2(-\q)$ in the PDC case.  
The correlation function $\langle
a^\dagger (\x) a(\xp) \rangle $ governs the properties of spatial
coherence of the thermal source \cite{bib10,Svelto}.  The correlation
length, or transverse coherence length $\lcoh$, is determined by the
inverse of the bandwidth $\Delta q$ of the function $\langle n(\q)
\rangle_{\rm th}$.  The same comments hold for the correlation
$\langle b_1 (\x) b_2(\xp) \rangle $, and the function
$U_1(\q)V_2(-\q)$ in the entangled case.  Most importantly, unlike the
results for the separable states considered in Refs. \cite{bib5,bib7},
in both Eqs.~(\ref{eq12}) and (\ref{eq9}) the modulus is outside the
integral; this is the feature that ensures the possibility of coherent
imaging via the correlation function (see, e.g., \cite{bib5}).

\section{Imaging schemes: image and diffraction pattern of an object}
Let us now analyse two paradigmatic examples of imaging systems,
borrowed from the discussion of \cite{bib7} and sketched in
Fig.~\ref{fig3}.
\begin{figure}[h]
{\scalebox{.4}{\includegraphics*{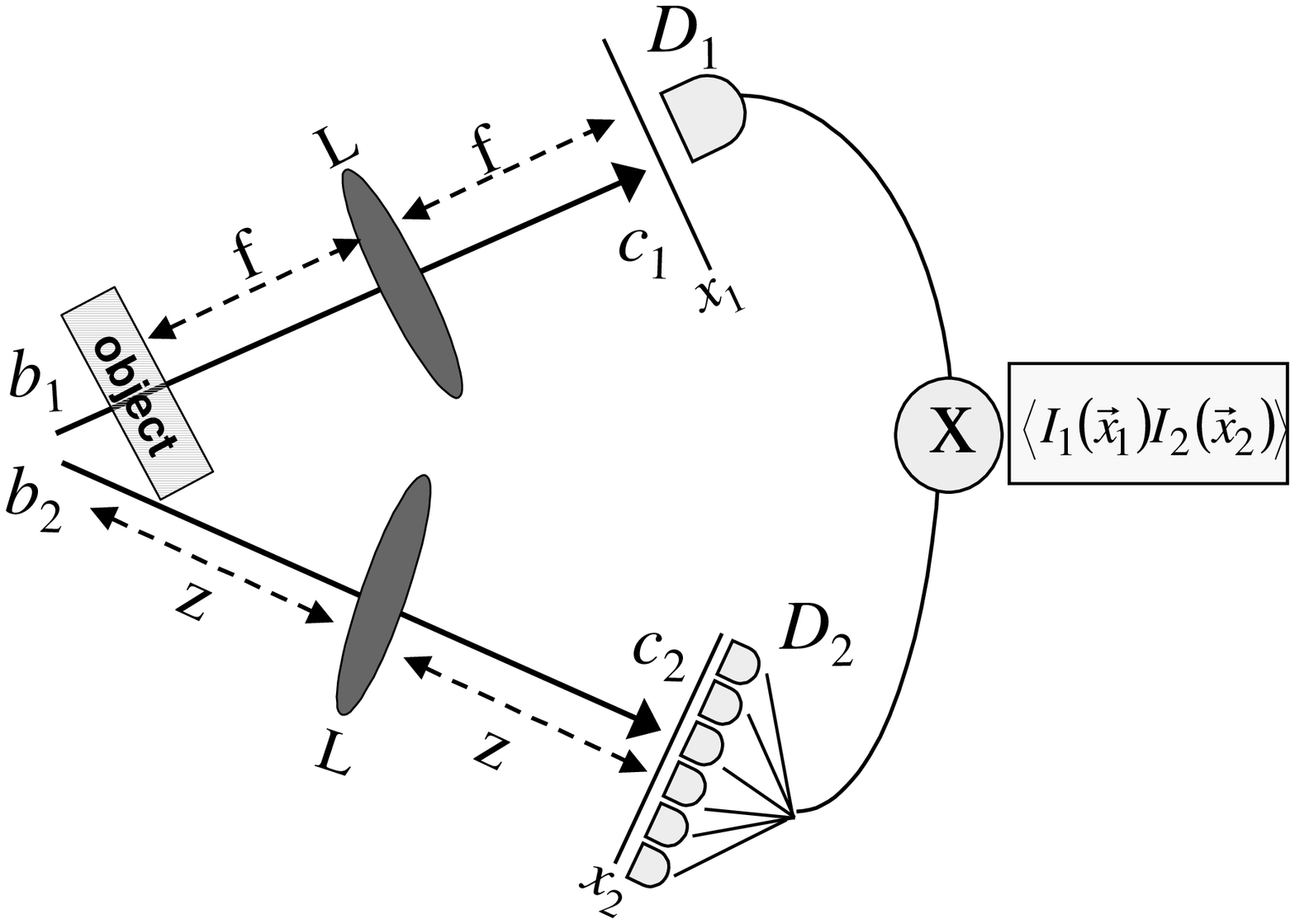}}} 
\caption{Imaging scheme. $L$ denotes two identical lenses of
focal length $f$. $D_1$ is a point-like detector.The distance $z$ is either $z=f$ or
$z=2f$.} \label{fig3}
\end{figure}
In both examples the setup of arm 1 is fixed, and consists of an
object, described by a complex transmission function $T(\x)$, and a
lens located at a focal distance $f$ from the object and from the
detection plane. Hence,
\begin{equation}
 h_1(\x_1,\xp_1)= -\frac{\im}{\lambda f} \exp{\left(-\frac{2\pi\im}{\lambda f} \x_1 \cdot \xp_1 \right)} T(\xp_1) \; ,
\label{h1}
\end{equation}
with $\lambda$ being the wavelength. In arm 2 there is a single lens
placed at a distance $z$ both from the source and from the detection
plane 2; for simplicity we take the two lenses identical. 

In the first example we assume $z=f$ so that $h_2(\x_2,\xp_2)=
-\frac{\im}{\lambda f} \exp{\left(- \frac{2\pi\im}{\lambda f} \x_2
    \cdot \xp_2\right) } \; .$ By inserting these propagators into
Eq.~(\ref{eq12}) and taking into account
Eq.~(\ref{gamma}), we obtain
\begin{equation}
G_{\rm th}(\x_1, \x_2) 
\propto \left| \langle n ( -\x_2 \frac{2\pi}{\lambda f} ) \rangle_{\rm th}  \, 
\tilde T \left( (\x_2 -\x_1) \frac{2\pi}{\lambda f}  \right)\right|^2 \: ,
\label{diffpatt}
\end{equation}
where $ \tilde T (\q) = \int \frac{\rm d \x}{2\pi} e^{-\im \q \cdot
  \x} T(\x)$ is the amplitude of the diffraction pattern from the
object. This has to be compared with the result of the entangled case
(see Eq.~(7) of \cite{bib7}),
\begin{equation}
G_{\rm PDC}(\x_1, \x_2) 
\propto \left| U_1 ( -\x_2 \frac{2\pi}{\lambda f} ) V_2 ( \x_2 \frac{2\pi}{\lambda f} )   \, 
\tilde T \left( (\x_2 +\x_1) \frac{2\pi}{\lambda f}  \right)\right|^2 \: ,
\label{diffpatt_ent}
\end{equation}
where the combination $\x_2+ \x_1$ appears instead of $\x_2 -\x_1$ and
$ U_1 V_2 $ instead of $\langle n \rangle_{\rm th}$. In both the thermal
and the PDC case the whole diffraction pattern from the object can be
reconstructed via the correlation function. This holds provided that the
spatial bandwidth $\Delta q$ is larger than the maximal transverse
wave-number $q$ in the diffraction pattern, or equivalently, provided
that $\lcoh < l_{\rm o}$, where $l_{\rm o}$ is the smallest scale of
variation of the object spatial distribution. Thus both cases
have best performances of the scheme when spatially incoherent light
($\lcoh \to 0$) is used. In contrast, as it is well known, when $\lcoh < l_{\rm o}$ no
information about the diffraction pattern of the object can be
obtained without the correlations, i.e. if we detect the light
intensity distribution in arm 1 with an array of pixels. In fact, one
can easily obtain that
\begin{equation}
\langle I_1(\x_1)\rangle  =|r|^2\int \frac{\rm d \q}{(\lambda f)^2}
\left|\tilde T(\x_1 \frac{2\pi}{\lambda f}-\q) \right|^2 \langle
n(\q)\rangle_{\rm th}\;.
\label{I1bis}
\end{equation}
For $\lcoh < l_{\rm o}$, $\langle n(\q)\rangle_{\rm th}$ can be taken out of the integral, and the resulting expression does not
depend on $\x_1$ any more.\\
We incidentally remark that the result of Eq.(\ref{diffpatt}) differs from what one would obtain with a standard Hambury-Brown and Twiss 
scheme \cite{bib16}, where the object is placed in the thermal beam $a$ before the beam-splitter.
In that case, one would retrieve the Fourier transform of the modulus square of the object transmission function (see e.g. \cite{bib10}), thus losing any phase information. In our scheme instead, where the object is located in only one arm of the two,  phase information about the object
can be extracted and e.g. the diffraction pattern from a pure phase object can be reconstructed. 

In the second example, we set $z=2f$, so that $h_2 (\x_2,\xp_2) =
\delta (\x_2+\xp_2) \exp{\left(-\im |\x_2|^2 \frac{\pi}{\lambda
      f}\right)}$.  Inserting this in Eq. (\ref{eq12}) and taking into
account (\ref{h1}), we get:
\begin{eqnarray}
G_{\rm th} (\x_1, \x_2) &\propto&
\left| \int {{\rm d}} \xp_1
 \Gamma (\xp_1 + \x_2) 
 T^* \left( \xp_1  \right) e^{\im \frac{2\pi}{\lambda f} \xp_1 \cdot \x_1}
\right|^2
\label{image1} \\
&\approx& 
\left| \langle n( \x_1 \frac{2\pi}{\lambda f} )   \rangle_{\rm th}\right|^2
  \left|  T \left( -\x_2  \right)
\right|^2 \: ,
\label{image2}
\end{eqnarray}
where in the second line $\lcoh < l_{\rm o}$ was assumed. Since the
correlation function $\Gamma (\xp_1 + \x_2)= \left\langle a^\dagger(
  \xp_1) a(-\x_2) \right\rangle$ is non-zero in a region of size
$\lcoh$ around $\xp_1=-\x_2$, this condition ensures that
$T(\xp_1)$ is roughly constant in this region and it can be taken out
from the convolution integral in (\ref{image1}), thus obtaining
(\ref{image2}).  In this example the intensity correlation function
provides information about the image of the object.  In the general
case~(\ref{image1}), the image reconstructed via the correlation function is a
convolution of the object image with the second order correlation
function (\ref{gamma}); therefore the thermal coherence length
$\lcoh$ fixes the resolution of the imaging scheme.

Under the same assumption $\lcoh < l_{\rm o}$, a similar result holds
for the case of entangled beams (see Eq.~(8) of
\cite{bib7})\footnote{There is unfortunately a misprint in Eq.~(8) of
  Ref. \cite{bib7}, so that the argument of the function $T$ is
  $\x_2$ instead of $-\x_2$. }
\begin{equation}
G_{\rm PDC}(\x_1,\x_2)
\propto
|T(-\x_2)|^2 \left|U_1 \left(\frac{2\pi\x_1 }{\lambda f} \right)V_2 \left(-\frac{2\pi\x_1}{\lambda f}
        \right)\right|^2 .
\label{image3}
\end{equation}
Also in this case the resolution of the scheme is limited by the
finite transverse coherence length of the PDC beams.

This section shows that the classical correlation of the thermal beams
offers imaging capabilities similar to those of the entangled PDC
beams; both the image and the diffraction pattern of an object can be
reconstructed and we can pass from one to the other by only operating
on the optical setup in the reference arm.  The performances of the
imaging schemes based on classical correlation and on entanglement
will be compared in Sec.~\ref{sec:Imag-perf-classical}, both for what
concerns the visibility, the statistical properties as well as the
spatial resolution.  The next section, instead, will explain the basic
mechanism that allows correlated thermal imaging with such a degree of
flexibility.

\section{Near and far-field correlation in the thermal and entangled case}
\label{comparison}
As explained in detail in Ref.~\cite{bib7} the imaging schemes
described in the previous section have a peculiar feature. In the
$z=f$ scheme the diffraction pattern reconstruction is made possible
by the presence of spatial correlation in the far field of the
correlated beams (momentum correlation of the photons). In the $z=2f$
scheme, it is the presence of spatial correlation in the near-field
(position correlation of the photons) that ensures the possibility of
reconstructing the image.  Our results for the thermal case may hence
appear surprising if one has in mind the case of a coherent beam
impinging on a beam splitter, where the two outgoing fields are
uncorrelated, i.e. $ G(\x_1, \x_2) =0$.  However, when the input field
is an intense thermal beam, i.e. the photon number per mode is not too
small, the two outgoing beams are well correlated in space both in
the near-field and in the far-field planes.

To prove this point, let us consider the number of photons detected in
two small identical portions $S$ (``pixels") of the thermal beams in
the near field immediately after the beam splitter, $N_i= \int_{S}
{\rm d} \x \, b_i^\dagger(\x) b_i (\x)\:$ $i=1,2$, and the difference
$N_-= N_1-N_2$. Making use of the transformation (\ref{BS}), for
$|r|^2= |t|^2=1/2$
it can be proven that the variance $\langle \delta N_-^2 \rangle =
\langle N_-^2\rangle -\langle N_-\rangle^2 $ is given by
\begin{equation}
\langle \delta N_-^2\rangle=
\langle N_1\rangle+ \langle  N_2\rangle\: ,
\label{SN}
\end{equation}
which corresponds exactly to the shot noise level. Remarkably,
Eq.~(\ref{SN}) holds regardless of the statistical properties of the
input beam $a$ provided that in the other input port there is the
vacuum.  On the other hand, by using the identity $\langle \delta
N_-^2\rangle= \langle \delta N_1^2\rangle + \langle \delta
N_2^2\rangle- 2\langle \delta N_1 \delta N_2 \rangle$ and taking into
account that $\langle \delta N_1^2\rangle=\langle \delta N_2^2\rangle$
for $ |r|^2=|t|^2$, the degree of spatial correlation is described by
\begin{equation}
C\stackrel{{\rm def}}{=}\frac{\langle \delta N_1 \delta N_2 \rangle }{\sqrt{\langle \delta N_1^2\rangle} 
\sqrt{ \langle \delta N_2^2\rangle} }
=1-\frac{\langle  N_1\rangle}{  \langle \delta N_1^2\rangle   } \; .
\label{C}
\end{equation}
For any state $0\le|C| \le 1$, where the upper bound is imposed by the
Cauchy-Schwarz inequality.  The lower bound corresponds to the
coherent state, for which $ \langle \delta N_1^2\rangle = \langle N_1
\rangle$ .  For the thermal state, there is always some excess noise
with respect to the coherent state $\langle \delta N_1^2\rangle >
\langle N_1 \rangle$, so that the correlation (\ref{C}) never
vanishes. Remarkably, a high degree of spatial correlation between
beams $b_1$ and $b_2$ is ensured by the presence of a high level of
excess noise in the input beam. As shown in detail in the Appendix,
for thermal systems with a large number of photons, provided that the
pixel size is on the order of $\lcoh$ or larger, $ \langle
N_1\rangle / \langle \delta N_1^2\rangle \ll 1$, and $C$ can be made
close to its maximum value. 
\par
Even more important, in the absence of losses it is not difficult to
show that Eqs.~(\ref{SN}) and (\ref{C}) hold in any plane linked to the
near-field plane by a Fresnel transformation.  Let us assume that the
propagation of beams $b_1,b_2$ is described by a linear and unitary
kernel $H$, $b_{H,i} (\x) = \int {\rm d} \xp H(\x, \xp) b_i (\xp) $,
$i=1,2$ . Then the form of the beam-splitter transformation
(\ref{BS}) is preserved during propagation, provided that the thermal
field $a$ is substituted by the propagated field $a_{H} (\x) = \int
{\rm d} \xp H(\x, \xp) a (\xp) $. Hence Eqs.~(\ref{SN}) and (\ref{C})
also hold for $b_{H,1}$ and $b_{H,2}$, because these equations are
just a consequence of the beam-splitter transformation (\ref{BS}) with
$|r|^2= |t|^2=1/2$. Moreover, the field $a_H$ after propagation is
still described by a thermal statistics, since the Gaussian statistics
and the factorization property (\ref{gamma2}) of the fourth-order
correlation function are preserved by a linear unitary transformation.
However, the coherence properties of the field $a_H$ change upon
propagation, as described by the well-known Van Cittert-Zernike
theorem (see, e.g., \cite{bib10}).  In particular, in the far-field
plane where $H (\x,\xp) =-\frac{\im}{\lambda f}
\exp{\left(-\frac{2\pi\im}{\lambda f} \x \cdot \xp \right)}$, they are
described by a second order correlation function
\begin{equation}
\langle a_H^\dagger (\x) a_H (\xp) \rangle \propto \int {\rm d} \vec{y} \int {\rm d} \vec{y}^{\, \prime}
e^{\im \frac{2\pi}{\lambda f}\left(\x \cdot \vec{y} - \xp \cdot \vec{y}^{\,\prime} \right) }
\Gamma \left( \vec y, \vec{y}^{\,\prime}  \right)   \; .
                        \label{VCZ}
\end{equation}
By assuming translational spatial invariance of the source, the
far-field correlation function (\ref{VCZ}) is proportional to
$\delta(\x-\xp)$, as can be easily verified by substituting
Eq.~(\ref{gamma}) into Eq.~(\ref{VCZ}). However, when this unrealistic assumption is removed, the
finite transverse size of the source $w_S$ has the effect that the
correlation length of the function (\ref{VCZ}) is also finite and is
inversely proportional to $w_S$, as shown by the Van Cittert-Zernike
theorem. Hence $l_{\rm coh}^{\, \prime} \propto \frac{\lambda f}{w_s}$,
represents the coherence length in the far-field plane.  We can thus
conclude that a high level of pixel-by-pixel correlation can be
observed also in the far-field plane, provided that the size of the
detection regions $S$ is not too small with respect to $l_{\rm coh}^{\,
  \prime}$, and the thermal beam is sufficiently intense (see the
discussion in the Appendix).

We remark that despite $C$ can be made
close to 1 by increasing the mean number of photons, it never reaches
the quantum level, as shown by Eq.~(\ref{SN}).

For the entangled beams produced by PDC, spatial correlation is
present both in the near and in the far field, with the ideal result
$\langle \delta N_-^2\rangle= 0$, $C=1$ in both planes \cite{bib8}. In
this case, the far field correlation is between symmetric pixels, and
the coherence length in the far field is inversely proportional to the
pump beam waist, which therefore in this context plays the same role as
the source size $w_S$ for the thermal beams. \\
 In \cite{bib7} we
analysed the effect of replacing the pure PDC entangled state with two
mixtures that exactly preserve the spatial signal-idler quantum
correlations, either in the far or in the near field. It turned out
that when considering the ``far-field mixture" the pure state
results could be exactly reproduced in the $z=f$ configuration of
Fig.~\ref{fig3}, but no information about the image was present in the
$z=2f$ configuration. The converse is true considering the
``near-field mixture".  This result is a consequence of the fact that
the far-field intensity and the near-field intensity are non-commuting
operators, and in the absence of quantum entanglement between the two
beams they cannot simultaneously be correlated up to a perfect degree
(this point is related to that raised in \cite{Boydarchive}).  This
led us to argue that only in the presence of quantum entanglement the
whole set of results can be obtained by solely changing the setup in
reference arm 2 \cite{bib7}.

However, nothing prevents two non-entangled beams to be correlated in
both planes up to an imperfect degree.  The two beams generated by
splitting thermal light are actually imperfectly correlated both in
the near and in the far-field; but by using intense thermal light, the
classical intensity correlation is strong enough to reproduce
qualitatively the results of both the $z=f$ and the $z=2f$
configuration. 

\section{Imaging performances in the classical and quantum regimes}
\label{sec:Imag-perf-classical}
A complete comparison of the performances in the classical and quantum
regimes requires extended numerical investigations
describing realistic thermal sources,
which are outside the scope of this paper.

However, some general remarks can be made concerning the key issue of
the visibility of the information in the two regimes. The information
about the object is retrieved by subtracting the background term $
\langle I_1 (\x_1) \rangle \langle I_2 (\x_2)\rangle $ from the
measured correlation function (\ref{eq5}), as indicated in
Eq.~(\ref{eq6}).  A measure of this visibility is given by evaluating
the following quantity in relevant positions
\begin{equation} 
{\cal V}  = \frac{G(\x_1,\x_2)}{\langle  I_1 (\x_1) I_2 (\x_2)\rangle } =
\frac{ G(\x_1,\x_2) }  {\langle  I_1 (\x_1) \rangle \langle I_2 (\x_2)\rangle + G(\x_1,\x_2) } \; ,
\label{visibility}
\end{equation}
with  $0 \le {\cal V}  \le 1$. 

A first remark concerns the presence of $\langle n(\q)\rangle_{\rm
  th}$ in Eq. (\ref{eq12}) in place of $ U_1(\q) V_2(-q)$ in
Eq.~(\ref{eq9}). As a consequence, in the thermal case $ G_{\rm
  th}(\x_1, \x_2)$ scales as $\langle n(\q)\rangle_{\rm th}^2$.  In
the entangled case, $ G_{\rm PDC}(\x_1, \x_2)$ scales as
$\left|U_1(\q) V_2(-q)\right|^2= \langle n(\q)\rangle_{\rm PDC} +
\langle n(\q)\rangle_{\rm PDC}^2 $, where $\langle n(\q)\rangle_{\rm
  PDC} =|V_2(-\q)|^2= |V_1(\q)|^2$ is the mean number of photons per
mode in the PDC beams, and $|U_1(\q)|^2 = 1+|V_1(\q)|^2$ (see, e.g.,
\cite{bib8}).  The difference between the two cases is immaterial when
the mean photon number is large, while it emerges clearly in the small
photon number regime ($ \langle n(\q)\rangle \ll 1$).  Actually, in
the thermal case the visibility does not exceed the value $1/2$,
whatever the value of $\langle n(\q)\rangle_{\rm th}$, since $ G_{\rm
  th}(\x_1, \x_2)$ scales in the same way as the background term.  On
the contrary, in the PDC case the visibility can approach the value
$1$ in the small photon number regime, since in this case the leading
scale of $G_{\rm PDC}(\x_1,\x_2)$ is $\langle n(\q)\rangle_{\rm PDC}$
and this term becomes dominant with respect to the background $\langle
I_1 (\x_1) \rangle \langle I_2 (\x_2)\rangle \propto \langle
n(\q)\rangle_{\rm PDC}^2$.  Hence, in the regime of single photon pair
detection the entangled case presents a much better visibility of the
information with respect to classically correlated thermal beams (see
also \cite{Klyshko}).

A second remark concerns the role of the temporal argument. Standard
calculations
show that the visibility scales as the ratio between the coherence
time of the source $\tau_{\rm coh}$ and the detection time (see also
\cite{bib8b,bib11}).
This implies that conventional thermal sources, with very small
coherence times, are not suitable for the schemes studied here. A
suitable source should present a relatively long coherence time, as
for example a sodium lamp, for which $\tau_{\rm coh} \approx 10^{-10}$~s
\cite{Svelto}, or the chaotic light produced by scattering a laser
beam through a random medium (see, e.g., \cite{bib15}). 

As a special example of a thermal source, one can consider the signal
field or the idler field generated by PDC.
Fig.~\ref{fig4} shows the results of a numerical simulation
\footnote{In our numerical simulations we dropped the translational
  invariance in both space and time of the input beam, and thus we
  also took into account the temporal variable (which was ignored in
  the analytical treatment). The simulations were done in 2D+1
  dimensions, i.e. including one transverse dimension (along the
  walk-off direction) as well as the time dimension, and propagating
  along the $z$-direction of the crystal. The calculated intensities
  were integrated over time since the pulse length was much shorter
  than the response time of any available detectors with spatial
  resolution. See \cite{bib8} for more details.}
\begin{figure}[h]
{\scalebox{.55}{\includegraphics*{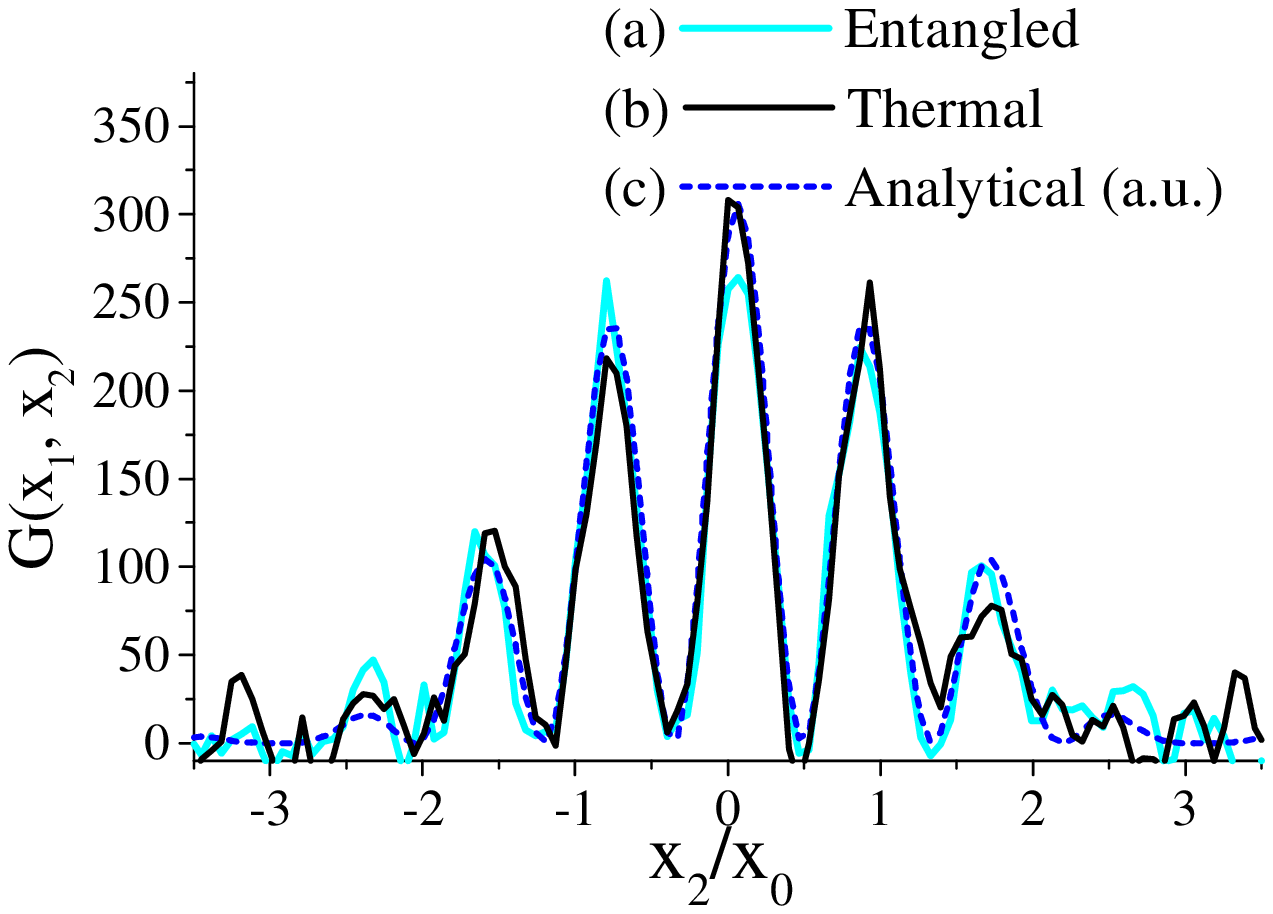}}} 
\caption{Numerical simulation of the reconstruction of 
  the diffraction pattern of a double slit in the scheme $z=f$ of
  Fig.~\ref{fig3}.  $G(\x_1,\x_2)$ is plotted versus $\x_2$ after
  $10^4$ pump shots for (a) entangled signal/idler beams from PDC, (b)
  classically correlated beams by splitting the signal beam. (c) is
  the analytical result of Eq.~(\ref{diffpatt}). Parameters are those of a 4 mm
  $\beta$-barium-borate crystal ($\lcoh= 16.6~ \mu$m, $\tau_{\rm
    coh}=0.97$~ps).  The pump waist is $664~ \mu$m, and the pulse
  duration is $1.5$~ps.  
The slits are $36
  ~\mu$m wide and slit separation is $122~\mu$m. $x_0$ is defined as
  $\Delta q \lambda f/(2\pi)$. } 
\label{fig4}
\end{figure}
for the reconstruction of the diffraction pattern of a double slit, in
the scheme $z=f$ of Fig.~\ref{fig3}. It compares the use of the
entangled signal and idler beams (curve (a)), and two classically
correlated beams obtained by symmetrically splitting the signal beam
(curve (b)). The parametric gain is such that $\langle n (\q)
\rangle_{\rm PDC} \approx 750$ at its maximum for (a), and $\langle n
(\q) \rangle_{\rm PDC} \approx 1500$ for (b), so that the the mean
photon numbers of beams $b_1$, $b_2$ are approximately equal in the
two simulations.  From our simulations it clearly emerged that the
number of pump shots necessary for reconstructing the diffraction
pattern up to a desired accuracy is the same for both curves (a) and
(b).  Notice that Fig.~\ref{fig4} plots only $G(\x_1,\x_2)$, which contains the object information and represents  the relevant part of the intensity correlation. This was obtained by subtracting a large background term $ \langle I_1 (\x_1) \rangle \langle I_2 (\x_2) \rangle$  from the correlation function of the intensities $ \langle I_1 (\x_1)  I_2 (\x_2) \rangle$. Consideration of these
quantities  
(not shown in the figure) allowed us to calculate the visibility ${\cal V}$,
as defined by Eq.(\ref{visibility}), which turned out to be ${\cal V}
\approx 0.05$ in both cases. Although this is a rather poor
visibility, the crucial point is that the fringes shown in Fig. \ref{fig4} could  be  correctly reconstructed
after a reasonable number of pump shots.

\begin{figure}[ht]
  {\scalebox{.55}{\includegraphics*{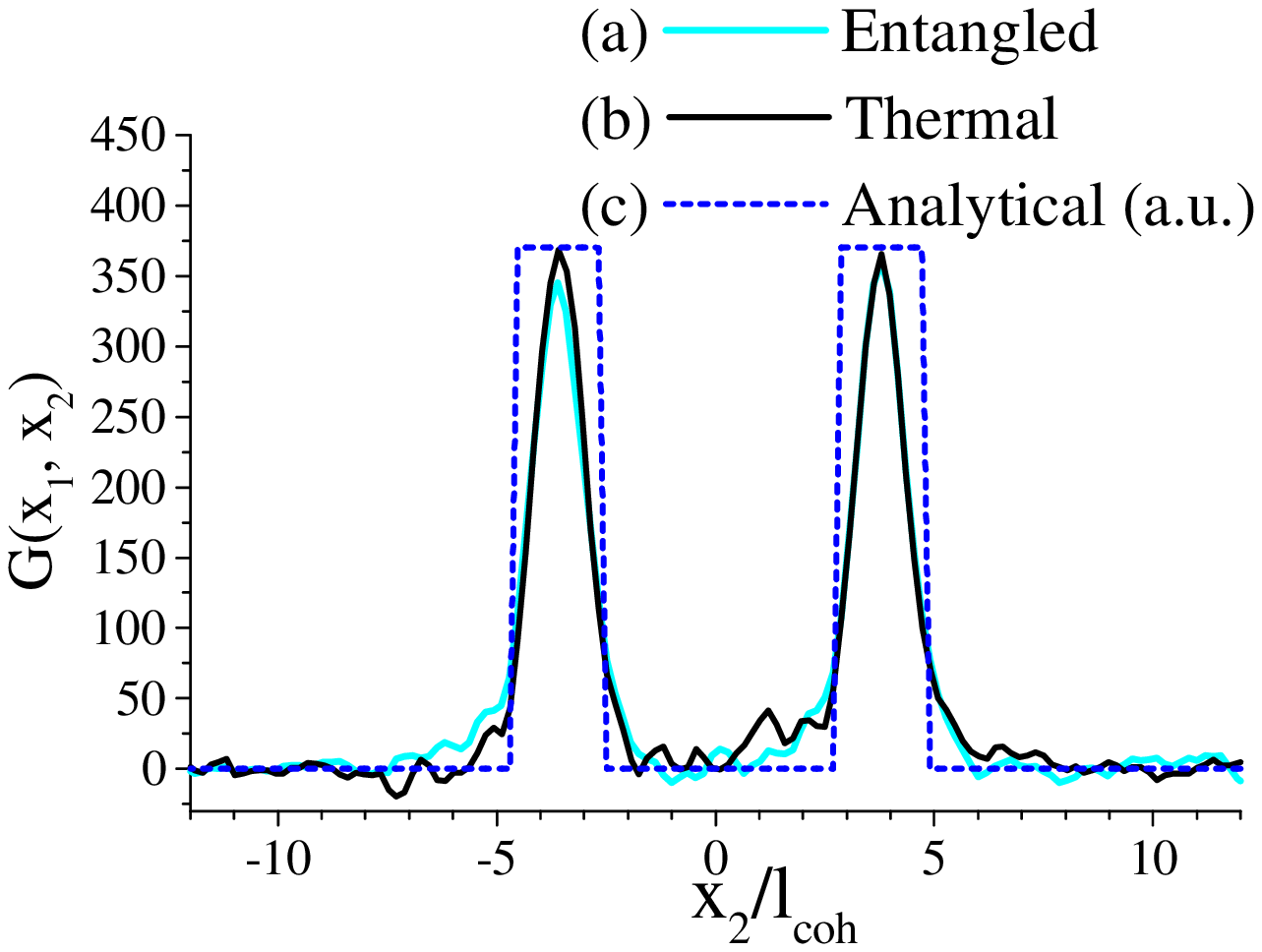}}} \caption{Numerical
    simulation of the reconstruction of the image of a double slit in
    the scheme $z=2f$ of Fig.~\ref{fig3}.  $G(\x_1,\x_2)$ is plotted
    versus $\x_2$ after $10^4$ shots for (a) entangled signal/idler
    beams from PDC, (b) classically correlated beams by splitting the
    idler beam. (c) is the analytical result of Eq.~(\ref{image2}).
    Parameters as in Fig.~\ref{fig4}.}
\label{fig5}
\end{figure}

The simulations of Fig.~\ref{fig4} were repeated but changing the setup in
the reference arm to the $z=2f$ configuration.  The results are shown
in Fig.~\ref{fig5}, and confirm that the classical correlation can be
used to reconstruct both the diffraction pattern and the image of the
object by operating only in the optical setup in the reference arm 2.
Also in this case the efficiency of the reconstruction is the same for
curves (a) and (b).

These examples clearly 
show that in the regime of high photon number the quantum and
classical correlations offer similar performances.

Another important aspect that emerges from these examples is that the
classical and quantum imaging schemes
apparently offer the same spatial resolution. This is most evident in
the plot of Fig.~\ref{fig5}, where the spatial resolution of both
schemes is not good enough to reproduce the sharp details of the
double slit image, but the reconstructed images are almost identical.
As a matter of fact the analytical results -- given by the general
formulas (\ref{eq12})-(\ref{eq9}) for the correlation function of
intensity fluctuations, the results
(\ref{diffpatt})-(\ref{diffpatt_ent}) for the diffraction pattern
reconstruction, and the results (\ref{image2})-(\ref{image3}) for the
image reconstruction -- show clearly that the spatial resolution of
both schemes depends only on the spatial coherence properties of the
sources.  Provided that the spatial coherence properties of the
classical source emulate those of the entangled source, there is no
reason why the two schemes should offer different spatial resolutions.
This requirement is not unrealistic at all, since the typical
transverse coherence length of the entangled beams from PDC is on the
order of tens of microns.
An example of
"thermal" light whose coherence properties can be engineered is
offered by, e.g., chaotic radiation obtained by scattering laser light
through random media \cite{bib15}.

A final remark concerns the form of the analytical results of Eqs.
(\ref{diffpatt}), (\ref{diffpatt_ent}) for the diffraction pattern
reconstruction. Specifically, the diffraction pattern on the r.h.s.
of Eq. (\ref{diffpatt_ent}) depends on the sum $\x_1 + \x_2$ of the
positions of the pixels in the detection planes of beam 1 and 2. This
feature was exploited in the experiment of Ref.~\cite{Shihlitho} (see
also \cite{giappo}), which was performed by registering
coincidence counts of pairs of photons generated by PDC in a
configuration similar to the $z=f$ scheme of Fig.~\ref{fig3}
\footnote{The scheme of the experiments of Ref.  \cite{Shihlitho} is
  slightly different from the $z=f$ scheme of Fig.~\ref{fig3}, and
  corresponds to having an object in both arms.  However, this only
  has the effect that the Fourier transform of the square of the
  object transmission function is observed instead of the Fourier
  transform of the object transmission function, which is irrelevant
  when the object is a double slit}.  The interference fringes from a
double slit
were observed by scanning the pixel detectors in the two beams
together, i.e. with $\x_1=\x_2$. A halving of the period of the
interference fringes with respect to those observed by illuminating
the object with coherent light was then observed. This effect was
claimed to be a consequence of the entanglement of the two-photon
state. This effect is evident from Eq. (\ref{diffpatt_ent}), where by
setting $\x_1+\x_2=\x$ the PDC correlation function gives $
\left|\tilde T \left( 2\x \frac{2\pi}{\lambda f} \right)\right|^2$.
By inspecting Eq.~(\ref{diffpatt}), we notice that the same effect
could in principle be observed in the scheme that uses the classically
correlated thermal beams, provided that the pixels in the detection
planes of the two beams are scanned symmetrically, i.e. setting
$\x_1=-\x_2$ \footnote{One can easily imagine a setup where in arm 1
  there is an additional optical element that flips the beam
  transverse distribution, so that $\x_1$ can be scanned together with
  $\x_2$ as in the experiments of \cite{Shihlitho} and
  \cite{giappo}.}.
Therefore, the discussion is open whether quantum entanglement has a
crucial role in this observed halving of the period of the interference
fringes, or whether this is just a consequence of the spatial
correlation of the two beams and of the particular detection scheme
used.

\section{Conclusions}
In conclusion we have suggested a  way of producing classically
correlated beams suitable for correlated imaging. These beams are
the outcome of mixing an intense thermal beam with the vacuum state on a
beam splitter. We have shown a deep analogy between the use of
entangled beams originating from PDC and these classically correlated
beams. The analogy arises because of the similar structure of Eq.
(\ref{eq12}) and Eq.~(\ref{eq9}), which implies that the
outcomes of correlation measurements with the classical source can
emulate those obtained with the entangled beams, for any choice of the
test and reference arm optical setups. This holds provided the
spatial coherence properties of the source are chosen to mimic the
marginal statistics of the individual PDC beams.

This analogy relies on the  high level of spatial correlation that exists
between the two beams emerging from the beam splitter as a consequence
of the large excess noise of the thermal input. A key point is that a
pixel by pixel correlation is present not only in the near-field plane
immediately after the beam splitter but also in the far-field plane.
Although the correlation is limited by shot noise, it allows to
reconstruct both the image and the diffraction pattern of an
object by only acting on the optical setup of the reference arm. This
was hitherto thought of as a feature that truly required entanglement
between the two beams \cite{bib7}.

 We have investigated the imaging performances of the quantum and classical schemes.
Both our analytical results and a specific numerical example show that
the spatial resolution limitations of the two schemes have similar origins, 
namely the finite transverse coherence
length of the light. Thus, also in this respect the classical beams
mimic the results of the entangled beams. 

On the other hand, in the small photon number regime a definite
advantage of the quantum configuration is represented by a better
visibility, as it was already recognized in other contexts (see, e.g.,
\cite{Franson}). Thus, in imaging schemes where the visibility
represents a crucial issue one should take this into account.
However, as the number of photons per mode becomes large this
advantage disappears and the visibility tends to be the same for the
quantum and the classical source. This result suggests that the peculiar
difference between the use of the two kind of sources is not given by
the entanglement, but rather by the possibility of working in the
photon counting regime in the quantum case. 

A further advantage of the PDC source may lie in the possibility of
using a fraction of the source laser as a reference field in order to
perform balanced homodyne detection. We will show in a future
theoretical work that the homodyne scheme makes it possible to perform
phase-sensitive correlated imaging with a high degree of visibility
even in the large photon number regime \cite{homodyne}. Another
advantage of using the PDC source relies on the possibility of
multi-wavelength imaging, as mentioned in the introduction.

Our results implies that it is possible to perform coherent imaging
without spatial coherence by using thermal light in combination with a
beam splitter. This is reminiscent of the Hanbury-Brown and Twiss
interferometric method for determining the stellar diameter
\cite{bib16}, as well as of detecting the fringes arising from interference
of two independent thermal sources \cite{bib17}.  However, here we
define a technique to achieve a fully coherent imaging of the object
through correlation measurements with a high degree of flexibility.
Moreover, since the required correlation is classical, a high quantum
efficiency of the detectors is not necessary.

The results of this paper suggest new experiments of correlated
imaging with thermal light, which could provide a useful test bed for
the imaging experiments with quantum entangled sources. 
Experiments like that simulated by the
numerics of Sec.~\ref{sec:Imag-perf-classical} may
open new possibilities, offered by the combination of the correlated imaging from entangled beams
and from classically correlated beams.

\acknowledgments{We are grateful to Alexander Sergienko, Bob Boyd, Ryan Bennink and Eric Lantz for stimulating discussions. This work was carried out in the framework of the FET
  project QUANTIM of the EU, of the PRIN project of MIUR "Theoretical study of novel devices based on quantum entanglement", and of the INTAS project "Non-classical light in quantum imaging and continuous variable quantum channels". M.B. acknowledges financial support from the Danish Technical
  Research Council (STVF).}
\section*{Appendix} 
This Appendix investigates how the degree of spatial correlation
between the beams obtained by splitting thermal light depends on the
size of the pixels used to detect the light, and on the mean number of
thermal photons.

As shown by Eq.~(\ref{C}) a high level of spatial correlation can be
present between the beams $b_1, b_2$ after the beam splitter as a
consequence of of a high level of excess noise in the thermal beam
$a$.  The relevant quantity to consider is the ratio $ \langle :
\delta N_1^2 :\rangle / \langle N_1 \rangle $, where $\langle : \delta
N_1^2 :\rangle $ is defined by $\langle \delta N_1^2 \rangle = \langle
N_1 \rangle + \langle : \delta N_1^2 :\rangle $, and represents the
noise in excess with respect to the coherent state level.  

By using the beam splitter transformation (\ref{BS}) and the
factorization property (\ref{gamma2}) of the fourth-order thermal
correlation function, we get
\begin{equation}
\langle : \delta N_1^2 :\rangle = |r|^4 \int_S {\rm d} \x \int_S {\rm d}\xp \left|\Gamma (\x- \xp) \right|^2
\: .
\label{EN} 
\end{equation}
where $\Gamma (\x-\xp)$ is the thermal correlation function defined by
Eq. (\ref{gamma}).  On the other hand, the mean number of photons
detected over the pixel is
\begin{equation}
\langle N_1 \rangle  = |r|^2 S \Gamma (0) \propto |r|^2 \frac{S}{l^2_{\rm coh}} n_{\rm max},
\: 
\label{N1} 
\end{equation}
where we introduced the parameter 
\begin{equation}
n_{\rm max}= \langle n(\q=0) \rangle_{\rm th}
\label{nmax}
\end{equation}
that represents the mean number of photon in the most intense mode
$\q=0$ in the spectrum. Moreover, we used $\Gamma(0)= \int \frac{{\rm
    d} \q}{(2\pi)^2} \langle n(\q) \rangle_{\rm th} \propto 1/l^2_{\rm coh}
n_{\rm max}$, where the proportionality constant depends on the actual
shape of the spectrum.  When the pixel size is much smaller than
$\lcoh$, the correlation function in Eq. (\ref{EN}) is approximately
constant over the integration regions $S$, so that
\begin{equation}
\langle : \delta N_1^2 :\rangle \to |r|^4 S^2 \Gamma (0)^2 = \langle N_1 \rangle ^2 
\: .
\label{ENsmall} 
\end{equation}
Hence, in this limit of a small detection region the noise takes the
form of the usual single-mode result for thermal light $\langle \delta
N_1^2 \rangle = \langle N_1 \rangle + \langle N_1 \rangle^2$. However,
as shown by Eq.~(\ref{N1}) the mean number of photons is also small in
this limit, so that the excess noise and the correlation tend to be
both small.

On the other side, when the detection regions grow much larger than
the coherence area the excess noise does not scale any more with the
square of the mean number of detected photons. In the limit of large
detection regions the r.h.s. of Eq. (\ref{EN}) can be approximated as
\begin{eqnarray}
\langle : \delta N_1^2 :\rangle &\to&   |r|^4 S  \int_{trv. plane}{\rm d } \vec{\xi} \left|\Gamma (\vec{\xi}) \right|^2
= |r|^4 S  \, \int \frac{{\rm d} \q}{(2\pi)^2}  \langle n (\q) \rangle^2_{\rm th} \nonumber\\
&\propto&  |r|^4 \frac{S}{l^2_{\rm coh}} n_{\rm max}^2 
\approx  |r|^2 n_{\rm max} \, \langle N_1 \rangle 
\: .
\label{ENlarge} 
\end{eqnarray}
For a large detection region the excess-noise scales proportionally to the mean
number of detected photons, so that the ratio $ \langle :
\delta N_1^2 :\rangle / \langle N_1 \rangle $ reaches a limiting value, which depends on
$n_{max}$. \\
We have investigated more in detail the role of the size of the detection region in
the degree of correlation  by assuming square pixels with sides
of size $\Delta$, as well as a Gaussian correlation function $\Gamma
(\vec{\xi} ) = \Gamma (0)
e^{-\frac{1}{2}|\vec{\xi}|^2/l^2_{\rm coh}}$, which is a good
approximation of a smooth correlation function decaying on a distance
$\lcoh$. By substituting this in Eq.  (\ref{EN}) and using the
result of Eq.~(\ref{C}), we get the two relevant asymptotic behaviours
of the degree of correlation (\ref{C}):
\begin{eqnarray}
C &\to& \frac{1}{4 \pi} \frac{\delta^2 n_{\rm max}} {1+ \frac{1}{4 \pi} \delta^2 n_{\rm max} }  \qquad {\rm for} \; \delta \ll 1
\\
C &\to& \frac{1}{4} \frac{ n_{\rm max}} {1+ \frac{1}{4} n_{\rm max} }  \qquad {\rm for} \; \delta \to \infty ,
\end{eqnarray}
where $\delta= \Delta/ \lcoh$ is the ratio of the pixel size to the
coherence length. The curves in Figure \ref{fig6} show the general
behaviour of the degree of correlation $C$ as a function of the pixel
size for different values of the mean photon number per mode $n_{\rm
  max}$.  The consequence of this figure is that when the input
thermal beam is intense enough, a high degree of spatial correlation
can be achieved.  This is provided that the pixel size is not too small with
respect to the coherence length describing the decay of the thermal
spatial correlation function.
\begin{figure}[h]
{\scalebox{.4}{\includegraphics*{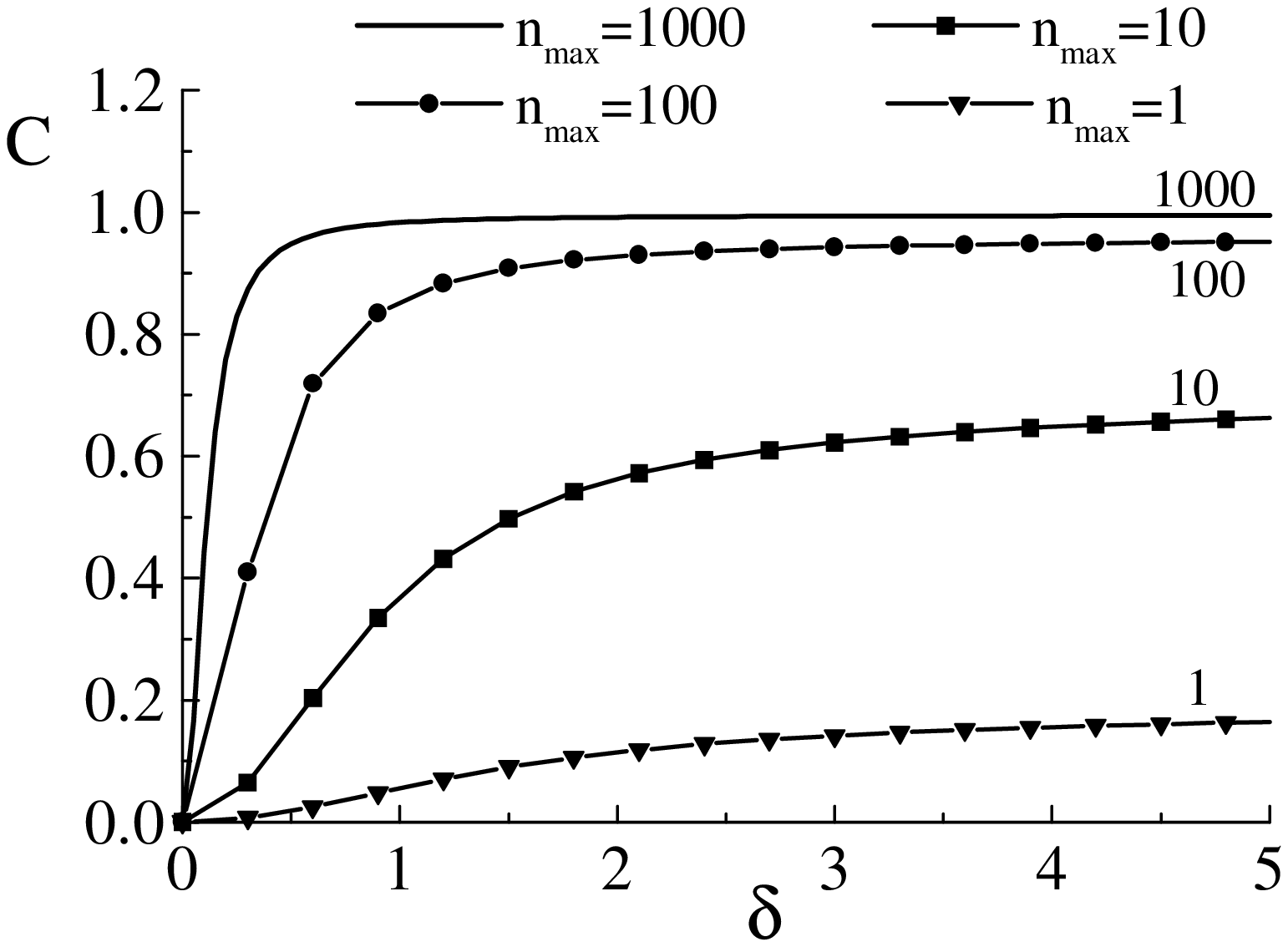}}}
\caption{Degree of spatial correlation $C$ between two identical
  detection regions of the beams obtained by splitting 
  thermal light, as a function of the ratio $\delta$ between the pixel
  size and the coherence length. $n_{\rm max}$ is the mean photon number
  in the most intense mode. $C=1$ represents the maximum degree of
  correlation.}
\label{fig6}
\end{figure}

Another relevant parameter that characterizes the spatial correlation
between two beams is the visibility of the correlation, which in the
spirit of Eq.~(\ref{C}) we define here as
\begin{equation}
  \label{eq:visi}
{\cal V}_S=\frac{\langle \delta N_1 \delta N_2\rangle}
{\langle  N_1 N_2\rangle }.
\end{equation}
This definition is analogous to that of Eq.(\ref{visibility}), where,
however, a small pixel was implicitly considered so that no integration over the pixel area
was performed.
This quantity is easily calculated in the same manner as above.
Figure~\ref{fig7} plots the visibility of the spatial correlation
between the thermal beams as a function of the pixel size scaled to
the coherence length.  This plot was obtained under the same
assumption of Fig.~\ref{fig6}. The visibility turns out not to depend
on the mean thermal photon number $n_{\rm max}$, so that only one
curve is plotted.
\begin{figure}[h]
{\scalebox{.4}{\includegraphics*{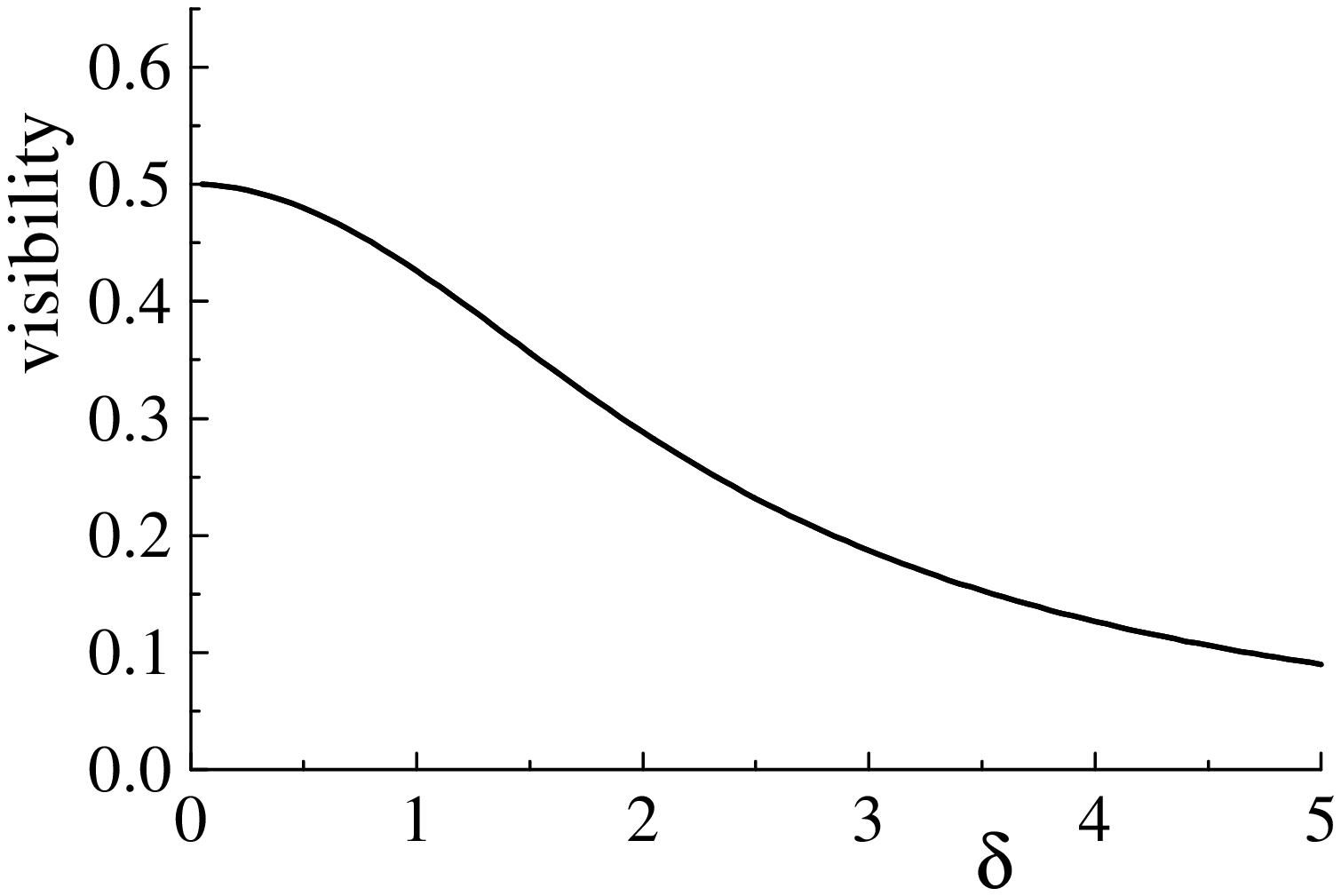}}}
\caption{Visibility ${\cal V}_S$
  of the spatial correlation between two identical detection regions
  of the beams obtained by splitting thermal light, as a function of
  the ratio $\delta$ between the pixel size and the coherence length.}
\label{fig7}
\end{figure}

Unlike the degree of correlation, the best visibility is obtained when
the detection pixel is not large with respect to the coherence length.
The main conclusion of this Appendix is therefore: in order to achieve
both a good visibility as well as a high degree of spatial correlation, the
best choice is a detection pixel with a size approximately equal to
the coherence area.

\end{document}